\documentclass[12pt,preprint,numberedappendix]{emulateapj}

\usepackage{courier,graphicx,epsfig,color,colordvi,natbib}
\usepackage{amsmath,bbold,mathrsfs}
\usepackage[english]{babel}
\usepackage{float}

\DeclareMathAlphabet{\mathscrbf}{OMS}{mdugm}{b}{n}

\begin{document}
\title{DELO-Bezier formal solutions of the polarized radiative transfer equation}
\author{J. de la Cruz Rodr\'iguez and N. Piskunov}
\affil{Department of Physics and Astronomy, Uppsala University, Box 516, SE-75120 Uppsala, Sweden}
\newcommand {\sv} {$\mathbf{S}$}
\newcommand {\kv} {$\mathbf{K}$}
\newcommand {\iv} {$\mathbf{I}$}
\newcommand {\ident} {$\mathbb{1}$}
\newcommand {\jv} {$\mathbf{j}$}
\def\mathbi#1{\textbf{\em #1}}

\begin{abstract}

We present two new accurate and efficient method to compute the formal
solution of the polarized radiative transfer equation. In this work, the source
function and the absorption matrix are approximated using quadratic
and cubic Bezier spline interpolants. These schemes provide
2$^\text{nd}$ and 3$^\text{rd}$ order approximation respectively and
don't suffer from erratic behavior of the polynomial approximation
(overshooting). The accuracy and the convergence of the new method are studied along with other popular
solutions of the radiative transfer equation, using stellar
atmospheres with strong gradients in the line-of-sight velocity and in
the magnetic-field vector.
\end{abstract}

\section{Introduction}\label{sec:intro}
During the past decade, large scale numerical models and inverse
problem applications became powerful tools for diagnosing and understanding
spectroscopic and spectropolarimetric observations.
Massive applications like 3D hydrodynamic (HD) simulations, Magnetic Doppler Imaging
and data inversion of solar surface layers stimulated interest in developing
fast and robust formal solvers for the radiative transfer equation (RTE).
These computationally demanding problems pose special requirements for the
RTE solver: a large number of integrations needed to compute
synthetic profiles per single iterations (time step or model adjustment).
Fast convergence become particularly important as it allows achieving acceptable
accuracy for the radiation field using the same geometrical grid as used for
hydrodynamics or inversion.

The properties of RTE solver become even more critical for non-equilibrium
modeling. The assumption of local thermodynamic equilibrium (LTE) makes
level population densities defined by the local temperature of the model
and decoupled from the radiation field. Only one integration of the RTE
is needed to compute the emerging intensity at each wavelength. However, in
the non-LTE case (NLTE), evaluation of population densities is
calculated using a self-consistent iterative method, normally requiring two
integrations of the RTE per iteration \citep{1987olson}. Inaccuracies in
the integration of the RTE on a coarser grid can lead to a slower convergence
of the NLTE problem in the spectral synthesis and generally to a
slower convergence of the inversion (in LTE and NLTE). The situation is
worse when the magnetic field becomes important.

The monochromatic radiative transfer equation for polarized light can be
expressed as:
\begin{equation}
  \frac{d\mathbi{I}}{ds}  = -\mathbf{K}\mathbi{I}+ \mathbi{j}\label{eq:rt}
\end{equation}
where $\mathbi{I}=(I,Q,U,V)^T$ is the Stokes parameter vector,
$\mathbi{j}=(j_I,j_Q,j_U,j_V)^T$ \ is
the total emission vector, and \kv \ is the total absorption matrix.
This matrix has seven independent terms: $\eta_I$ is the total
absorption of radiation; $\eta_Q$,
$\eta_U$ and $\eta_V$ describe the coupling of the intensity $I$
with $Q$, $U$ and $V$; and $\rho_Q$, $\rho_U$ and $\rho_V$ are
cross-talk terms between $Q$, $U$ and $V$ due to magneto-optical
effects \citep[see][]{2004egidio}:
\begin{equation}
  \mathbf{K}= \begin{pmatrix}
      \eta_I & \eta_Q & \eta_U & \eta_V \\[0.3em]
      \eta_Q & \eta_I & \rho_V & -\rho_U \\[0.3em]
      \eta_U & -\rho_V & \eta_I & \rho_Q \\[0.3em]
      \eta_V & \rho_U & -\rho_Q & \eta_I \\[0.3em]
      \end{pmatrix}.
\end{equation}

A number of advanced integration schemes, suitable for HD simulations
and NLTE inversions have been implemented for the unpolarized light
\citep{2003auer}, however, for polarized light it is common to
use a short characteristics schemes assuming linear \citep{1989rees}
or parabolic
dependence \citep{2003trujillo-bueno,2008sampoorna} of the source function with
optical path.

\citet{1998bellot-rubio} proposed an efficient third-order method (LBR hereafter)
that provides accurate results on coarse grids of depth-points for
polarized light. This method has been used extensively to compute LTE
inversions  \citep{2000bellot-rubio,2011socas-navarro} and NLTE inversions \citep{2012delacruz}.

In this paper, we propose a new method to accurately integrate the
polarized RTE, using Bezier interpolants. The paper is arranged as follows: an introduction to the DELO method and Bezier Splines 
are provided in \S~\ref{sec:DELO} and \ref{sec:bez} respectively. For completeness, we introduce the solution to the 
unpolarized RTE, using Bezier interpolants, in \S~\ref{sec:bezscal}. The quadratic and cubic Bezier solutions
to the polarized RTE are presented, for the first time, in \S~\ref{sec:bezpol}. Our proposed solutions are tested
against other methods commonly used in radiative transfer studies in \S~\ref{sec:numcalc}. Finally, our conclusions
are summarized in \S~\ref{sec:discussion}.

\section{The DELO method}\label{sec:DELO}

\cite{1989rees} rewrite the polarized RTE using the
\emph{modified source vector}
($\mathbi{S}=\mathbi{j} / \eta_I$), dividing all terms in
Equation~(\ref{eq:rt}) by the absorption coefficient $\eta_I$.  This simple
change leaves the \kv \ matrix with all the diagonal elements
normalized to one so the method got a name Diagonal Element Lambda
Operator (DELO).

Defining,
\begin{equation}
\mathbf{\bar{K}}=\frac{\mathbf{K}}{\eta_I} - \mathbb{1},
\end{equation}
where $\mathbb{1}$ is the $4\times4$ identity matrix. The radiative transfer
equation becomes:
\begin{equation}
  \frac{d\mathbi{I}}{d \tau_\nu} = \mathbi{I} -\mathscrbf{S},\label{eq:DELO}
\end{equation}
with the optical path defined as $d\tau_\nu=\eta_I ds$ and
$\mathscrbf{S} = \mathbi{S} - \mathbf{\bar{K}}\mathbi{I}$.

In our discrete grid of $n$ depth points $(k=1,2,3,...,n)$, the
solution to Equation~(\ref{eq:DELO}) on the interval $(\tau_k,\tau_{k+1})$ is:
\begin{equation}
  \mathbi{I}(\tau_k) = \mathbi{I}(\tau_{k+1}) e^{-\delta_k} + \int_{\tau_{k}}^{\tau_{k+1}} e^{-(\tau-\tau_k)}\mathscrbf{S}(\tau)d\tau,\label{eq:DELOsol}
\end{equation}
where $\delta_k\equiv d\tau_k$.

To integrate analytically Equation~(\ref{eq:DELOsol}), \cite{1989rees}
assume that the equivalent source vector $\mathscrbf{S}$ is linear with
optical depth (conventionally referred to as DELO-linear). A quadratic
dependence of $\mathscrbf{S}$ with optical depth would ideally allow to
improve the convergence, however, it becomes unstable in
the presence of non-linear gradients \citep{1990murphy}.

\section{The Bezier interpolants}\label{sec:bez}
\cite{2003auer} provides a thorough summary of advanced interpolants
that could be used to integrate the (implicitly assumed) unpolarized
radiative transfer equation.

\subsection{Bezier quadratic interpolant}\label{sec:qbez}
Defining normalized abscissa units in the interval $(x_k,x_{k+1})$,
\begin{equation}
 u=\frac{x-x_k}{x_{k+1} - x_k},
\end{equation}
the quadratic Bezier interpolant can be expressed as:
\begin{equation}
  f(x) = (1-u)^2y_k + y_{k+1}u^2 + 2u(1-u)\cdot C, \label{eq:bezier}
\end{equation}
where $y_k$ and $y_{k+1}$ represent the node values of the
function that is being interpolated and $C$ is the Bezier control
point. The latter can be expressed using the derivative at $x_{k}$ or
$x_{k+1}$,
\begin{eqnarray}
  C^0 (x_k)     &=& y_k     + \frac{h_k}{2}y'(x_k),     \label{eq:con0}\\
  C^1 (x_{k+1}) &=& y_{k+1} -  \frac{h_k}{2}y'(x_{k+1}), \label{eq:con1}
\end{eqnarray}
with $h_k=x_{k+1}-x_k$. If both $C^0$ and $C^1$ can be computed, it is
desirable to take the mean:
\begin{equation}
C =\left(C^0+C^1\right)/2, \label{eq:tcon}
\end{equation}
resulting in a more "symmetric" spline.

Defining,
\begin{eqnarray}
  \alpha = \frac{1}{3}\left(1+\frac{h_k}{h_k+h_{k-1}}\right),\label{eq:weight}\\
  d_{k+1/2} = \frac{y_{k+1} - y_k}    {h_k}, \label{eq:d+}\\
  d_{k-1/2} = \frac{y_{k}   - y_{k-1}}{h_{k-1}},     \label{eq:d-}
\end{eqnarray}
accurate numerical derivatives are calculated at the node points $(x_{1,2,...,n})$ with
a scheme that suppresses overshooting beyond the node function values \citep{1984Fritsch}.
The Bezier derivative is given by:
\begin{equation}
y'(x_i) =
    \frac{d_{i-1/2}\cdot d_{i+1/2}}{\alpha\cdot d_{i+1/2} + (1-\alpha)\cdot d_{i-1/2}},
\label{eq:deriv}
\end{equation}
if $d_{k-1/2}\cdot d_{k+1/2}>0$ and it is set to 0 otherwise.

\subsection{Bezier cubic interpolant}\label{sec:cbez}
Alternatively, the cubic Bezier interpolant can be expressed as:
\begin{equation}
\begin{split}
  f(x) =& (1-u)^3y_k + y_{k+1}u^3 + \\
 &3u(1-u)^2 \cdot \hat{E} +  3u^2(1-u) \cdot \hat{F}, \label{eq:bezier3}
\end{split}
\end{equation}
where $\hat{E}$ and $\hat{F}$ are control points, similar to those
defined in Equations~(\ref{eq:con0}) and (\ref{eq:con1}), but each of them is
placed in a different location:
\begin{eqnarray}
  \hat{E} (x_k)     &=& y_k     + \frac{h_k}{3}y'(x_k),     \label{eq:ccon0}\\
  \hat{F} (x_{k+1}) &=& y_{k+1} -  \frac{h_k}{3}y'(x_{k+1}). \label{eq:ccon1}
\end{eqnarray}

\begin{figure}
\resizebox{\hsize}{!}{\includegraphics[]{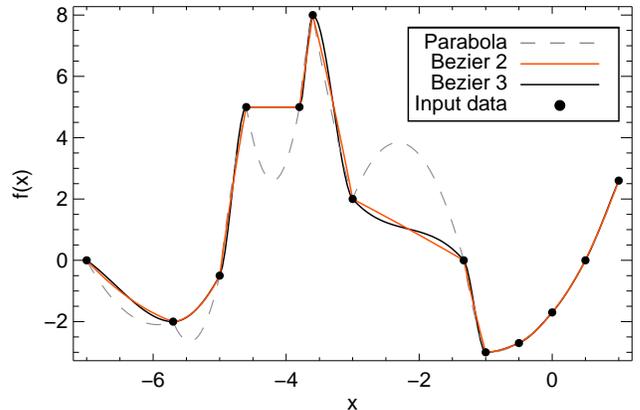}}
\caption{Comparison of 3 interpolation schemes. The black circles
  indicate the know values of the function that is interpolated using
  quadratic Bezier spline, cubic Bezier spline and piece-wise parabolic fit.}\label{fig:splines}
\end{figure}

Fig.~\ref{fig:splines} illustrates the behavior of three
interpolation schemes, when the function has fast changing
gradient. Here, we compare a
parabolic polynomial, the quadratic Bezier splines (Bezier 2) and the cubic
Bezier splines (Bezier 3) described in this section.

The behavior of the parabolic fit is erratic in the vicinity of intervals with
large change of gradient, showing dramatic overshooting peaks. The Bezier
splines considered here stays between the pair of boundary data values in each interval.
If fact, \citet{2003auer} already suggested that the Bezier splines (and the Hermitian
interpolant) are particularly suitable for solving the radiative transfer equation
in form (\ref{eq:DELOsol}) as they do not produce spurious maxima and minima.

From $x=-1$ to $x=1$ the curve increases smoothly, and the three
interpolation schemes considered here produce very similar values.

\section{Bezier integration of the scalar RTE}\label{sec:bezscal}
In this section, we derive the formal solution of the radiative transfer equation
for unpolarized light $(\mathbf{\bar{K}}=0)$. In this case Equation~(\ref{eq:DELO}) becomes
a scalar equation:
\begin{equation}
  \frac{dI}{d\tau} = I - S,
\end{equation}
where $I$ is the intensity and $S$ is the unpolarized source
function. Equation~(\ref{eq:DELOsol}) is easily integrated by approximating
$S$ with any of the Bezier interpolants described in
Sec.~\ref{sec:bez}.

\subsection{Quadratic Bezier integration}\label{sec:qbezscal}
If the quadratic Bezier interpolant is used to approximate the source
function, the solution to Equation~(\ref{eq:DELOsol}) is
\begin{equation}
I(\tau_{k}) = I(\tau_{k+1})\cdot e^{-\delta_k} + \alpha_k S_{k} + \beta_k
S_{k+1} + \gamma_k C_k, \label{eq:DELOscalsol}
\end{equation}
where $\alpha_k$, $\beta_k$ and $\gamma_k$ are coefficients from the
integral that only depend on $\delta_k\equiv \tau_{k+1}-\tau_k$:
\begin{eqnarray*}
  \alpha_k &=& \frac{2+\delta^2_k -2\delta_k -
    2e^{-\delta_k}}{\delta^2_k},\\
  \beta_k &=&
  \frac{2-(2+2\delta_k+\delta_k^2)e^{-\delta_k}}{\delta_k^2},\\
  \gamma_k &=& \frac{2\delta_k-4+(2\delta_k+4)e^{-\delta_k}}{\delta_k^2}.
\end{eqnarray*}

To calculate $\delta_k$, the opacity can be integrated analytically by using Bezier
  splines to approximate $\eta_I$ as a function of
  depth. Further details can be found in Appendix~\ref{sec:apen1}. 

Note, that for small $\delta_k$ it is wise to use Taylor expansion for
the exponents in the right-hand-side to avoid division of vanishingly
small quantities (see Appendix~\ref{sec:apen2}).

Equation~(\ref{eq:DELOscalsol}) has the form $I_k=A\cdot I_{k+1}+B$. In the case
of a stellar atmosphere we can sequentially evaluate outgoing intensity at
any depth point starting from the boundary condition $I_n=S_n$ at the bottom.
The control point $C_k$ is computed using Equation~(\ref{eq:tcon}) for all inner
points of the integration domain. For the top point only one approximation
for the control point ($C^1$) is available.

\subsection{Cubic Bezier integration}
If the cubic Bezier interpolant is used to approximate the source
function, the solution to Equation~(\ref{eq:DELOsol}) is
\begin{equation}
\begin{split}
I(\tau_{k}) =& I(\tau_{k+1})\cdot e^{-\delta_k} + \hat{\alpha}_k S_{k} +\hat{\beta}_k S_{k+1} \\
&+ \hat{\gamma}_k \hat{E}_k + \hat{\epsilon}_k \hat{F}_k. \label{eq:cDELOscalsol}
\end{split}
\end{equation}

$\hat{\alpha}_k$, $\hat{\beta}_k$, $\hat{\gamma}_k$, $\hat{\epsilon}_k$ are coefficients from the
integral that only depend on $\delta_k$:
\begin{eqnarray*}
\hat{\alpha}_k &=&  \frac{-6+6\delta_k-3\delta_k^2+\delta_k^3+6e^{-\delta_k}}{\delta_k^3},\\
\hat{\beta}_k &=& \frac{6+(-6-\delta_k(6+\delta_k(3+\delta_k)))\cdot e^{-\delta_k}}{\delta_k^3},\\
\hat{\gamma}_k &=&3 \cdot
\frac{6+(\delta_k-4)\delta_k-2(\delta_k+3)\cdot e^{-\delta_k}}{\delta_k^3},\\
\hat{\epsilon}_k &=&3 \cdot \frac{2\delta_k-6 + (6+\delta_k^2
    +4\delta_k)\cdot e^{-\delta_k} }{\delta_k^3}.
\end{eqnarray*}
The same recommendations apply here for small $\delta_k$.

\section{Bezier Integration of the vector RTE}\label{sec:bezpol}
\subsection{Quadratic Bezier integration}
For the polarized case $(\mathbf{\bar{K}}\neq 0)$, the solution to
Equation~(\ref{eq:DELOsol}) can formally be presented in the form similar to Equation~(\ref{eq:DELOscalsol}):
\begin{eqnarray}
  \mathbi{I}(\tau_k) &=& \mathbi{I}(\tau_{k+1})\cdot e^{-\delta_k} + \alpha_k\mathscrbf{S}(\tau_{k})+ \nonumber \\
  &+& \beta_k \mathscrbf{S}(\tau_{k+1}) + \gamma_k \mathbi{C}_k. \label{eq:DELOlam}
\end{eqnarray}

In Sec.~\ref{sec:bezscal}, Equation~(\ref{eq:deriv}) we show how to compute the
derivatives of $S$ needed for evaluating the control point $C$.
In the polarized case, however, derivatives of $\mathbi{I}$,
$\mathbi{S}$ and $\mathbf{\bar{K}}$ must be computed. The presence of $\mathbi{I}_k$ in the denominators of both $C^0$ and $C^1$
(through Equation~(\ref{eq:deriv})) introduces non-linearity that kills a simple
recurrence relation available in the scalar case:
$\mathbi{I}_k=\mathbf{A}\cdot \mathbi{I}_{k+1}+\mathbi{B}$.

An elegant solution to this problem is actually provided by the radiative
transfer equation:
\begin{eqnarray}
\mathbi{I}'_k&=&\mathbi{I}_k -\mathbi{S}_k + \mathbf{\bar{K}}_k
\mathbi{I}_k,\\
\mathbi{I}'_{k+1}&=&\mathbi{I}_{k+1} -\mathbi{S}_{k+1} + \mathbf{\bar{K}}_{k+1}
\mathbi{I}_{k+1}.
\end{eqnarray}
Using these expressions, the control points can be re-written as:
\begin{equation}
\begin{split}
   \mathbi{C}^0_k=&\overbrace{\left(\mathbb{1} +
   \frac{\delta_k}{2}\mathbf{\bar{K}}_k\right)\mathbi{S}_k+\frac{\delta_k}{2}\mathbi{S}'_k}^{\mathbi{c}_k^0}-\\
  &\underbrace{-\left[\frac{\delta_k}{2}\left(\mathbf{\bar{K}}_k\mathbf{\bar{K}}_k+\mathbf{\bar{K}}'_k+\mathbf{\bar{K}}_k\right)
  +\mathbf{\bar{K}}_k\right]}_{\bar{\mathbf{c}}_k^0} \mathbi{I}_k=\\
  =& \ \mathbi{c}_k^0+\mathbf{\bar{c}}_k^0\mathbi{I}_k,
\end{split}
\end{equation}
and:
\begin{equation}
\begin{split}
   \mathbi{C}^1_k=&\left(\mathbb{1} -
   \frac{\delta_k}{2}\mathbf{\bar{K}}_{k+1}\right)\mathbi{S}_{k+1}-\frac{\delta_k}{2}\mathbi{S}'_{k+1}+\\
  +&\left[\frac{\delta_k}{2}\left(\mathbf{\bar{K}}_{k+1}\mathbf{\bar{K}}_{k+1}+\mathbf{\bar{K}}'_{k+1}+
   \mathbf{\bar{K}}_{k+1}\right)\right.-\\
  -&\left.\mathbf{\bar{K}}_{k+1}\right]\mathbi{I}_{k+1}.
\end{split}
\end{equation}
Note that $\mathbi{C}^1_k$ and $\mathbi{c}_k^0$ are vectors, whereas
$\mathbf{\bar{c}}_k^0$ is a matrix.

The derivatives of the modified opacity matrix $\mathbf{\bar{K}}'$ and the source
vector $\mathbi{S}'$ must be computed according to the recipe for monotonicity given by
Equations~(\ref{eq:weight})-(\ref{eq:deriv}) to prevent the overshooting.

This convenient splitting of $\mathbi{C}^0$ permits re-writing
Equation~(\ref{eq:DELOlam}) for unknown Stokes vector in point $k$ as a system of four linear equations $(\mathbf{A}\cdot\mathbi{I}_k~=~\mathbi{B})$:
\begin{equation}
  \underbrace{\left(\mathbb{1}+\alpha_k\mathbf{\bar{K}}_k-\frac{\gamma_k}{2}\mathbf{\bar{c}}_k^0\right)}_{\mathbf{A}}
\mathbi{I}_k=\underbrace{\boldsymbol \xi_k +\frac{\gamma_k}{2}(\mathbi{C}_k^1+\mathbi{c}_k^0)}_{\mathbi{B}},
\label{eq:polsol}
\end{equation}
with,
\begin{equation}
\boldsymbol \xi_k = \mathbi{I}_{k+1}e^{-\delta_k} +\alpha_k\mathbi{S}_k+
\beta_k\mathscrbf{S}_{k+1} .
\end{equation}

Note, that the matrix and the right-hand-side in Equation~\ref{eq:polsol}, include only known quantities of absorption matrix, source and Stokes vectors. Note also, that solving the system of linear equations (\ref{eq:polsol}) directly instead of trying to invert \textbf{A} is both faster and more robust.

\subsection{Cubic Bezier integration}
Similarly to the quadratic Bezier integration, the solution to
Equation~(\ref{eq:DELOsol}) is formally similar to Equation~(\ref{eq:cDELOscalsol}):
\begin{eqnarray}
  \mathbi{I}(\tau_k) &=& \mathbi{I}(\tau_{k+1})\cdot e^{-\delta_k} + \hat{\alpha}_k\mathscrbf{S}(\tau_{k})+ \nonumber \\
  &+& \hat{\beta}_k \mathscrbf{S}(\tau_{k+1}) + \hat{\gamma}_k \hat{\mathbi{E}}_k + \hat{\epsilon}_k \hat{\mathbi{F}}_k. \label{eq:DELOcvec}
\end{eqnarray}

Given the formal similarity of this equation with the quadratic case, where also two
control points are used, the algebra needed to write the problem as a linear system
of equations (Equation~(\ref{eq:polsol})) is almost identical.
\begin{equation}
\begin{split}
   \hat{\mathbi{E}}_k=&\overbrace{\left(\mathbb{1} +
   \frac{\delta_k}{3}\mathbf{\bar{K}}_k\right)\mathbi{S}_k+\frac{\delta_k}{3}\mathbi{S}'_k}^{\mathbi{e}_k}-\\
  &\underbrace{-\left[\frac{\delta_k}{3 }\left(\mathbf{\bar{K}}_k\mathbf{\bar{K}}_k+\mathbf{\bar{K}}'_k+\mathbf{\bar{K}}_k\right)
  +\mathbf{\bar{K}}_k\right]}_{\bar{\mathbf{e}}_k} \mathbi{I}_k=\\
  =& \ \mathbi{e}_k+\mathbf{\bar{e}}_k\mathbi{I}_k,
\end{split}
\end{equation}
and:
\begin{equation}
\begin{split}
   \hat{\mathbi{F}}_k=&\left(\mathbb{1} -
   \frac{\delta_k}{3}\mathbf{\bar{K}}_{k+1}\right)\mathbi{S}_{k+1}-\frac{\delta_k}{3}\mathbi{S}'_{k+1}+\\
  +&\left[\frac{\delta_k}{3}\left(\mathbf{\bar{K}}_{k+1}\mathbf{\bar{K}}_{k+1}+\mathbf{\bar{K}}'_{k+1}+
   \mathbf{\bar{K}}_{k+1}\right)\right.-\\
  -&\left.\mathbf{\bar{K}}_{k+1}\right]\mathbi{I}_{k+1}.
\end{split}
\end{equation}

Equation~(\ref{eq:DELOcvec}) can be re-arranged to group all the terms
containing $\mathbi{I}(\tau_k)$ on the left-hand side of the equation.
\begin{equation}
  \begin{split}
    &  \underbrace{\left(\mathbb{1}+\hat{\alpha}_k\mathbf{\bar{K}}_k-\hat{\gamma}_k\mathbf{\bar{e}}_k\right)}_{\mathbf{A}}
    \mathbi{I}_k=\\
    &=\underbrace{\mathbi{I}_{k+1}e^{-\delta_k} +\hat{\alpha}_k\mathbi{S}_k+
      \hat{\beta}_k\mathscrbf{S}_{k+1} +\hat{\gamma}_k \mathbi{e}_k+\hat{\epsilon}_k\hat{\mathbi{F}}_k}_{\mathbi{B}}.
    \label{eq:cpolsol}
  \end{split}
\end{equation}

Again we advise to solve the linear system of equations in
Equation~(\ref{eq:cpolsol}), instead of inverting the $\mathbi{A}$
matrix.

\begin{figure*}[!ht]
   \centering
   \resizebox{\hsize}{!}{\includegraphics[]{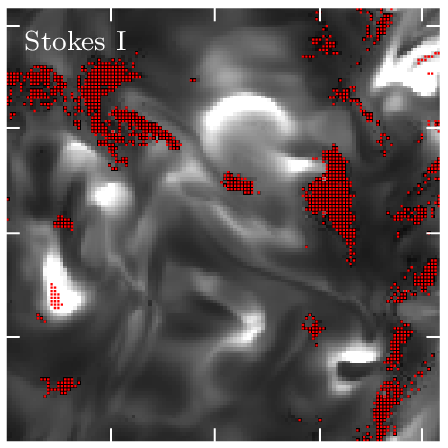}\includegraphics[]{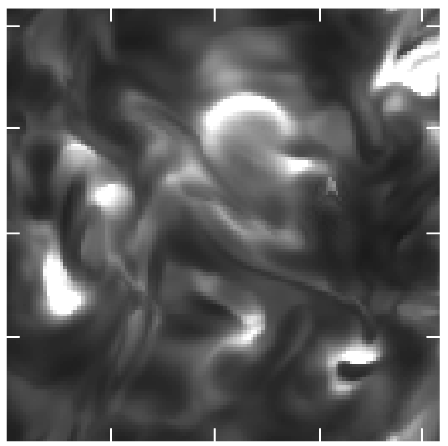}\includegraphics[]{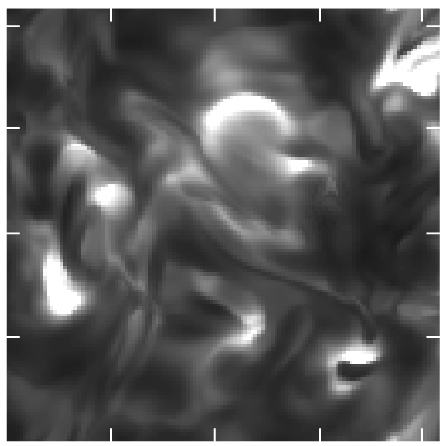}\includegraphics[]{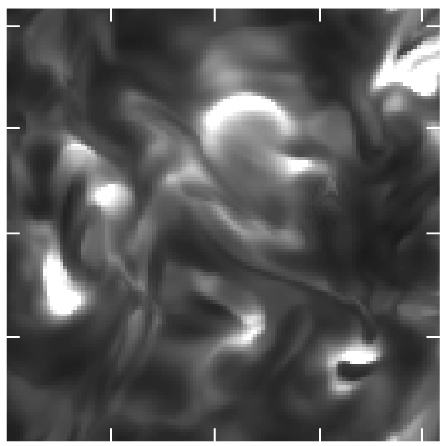}}
   \resizebox{\hsize}{!}{\includegraphics[]{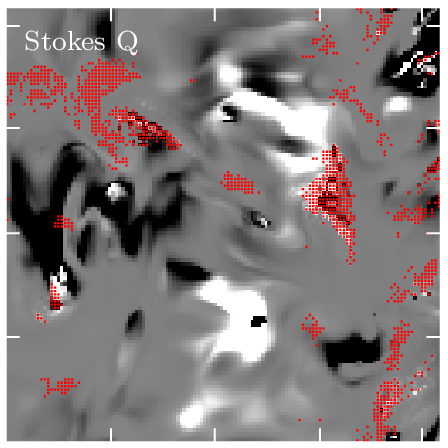}\includegraphics[]{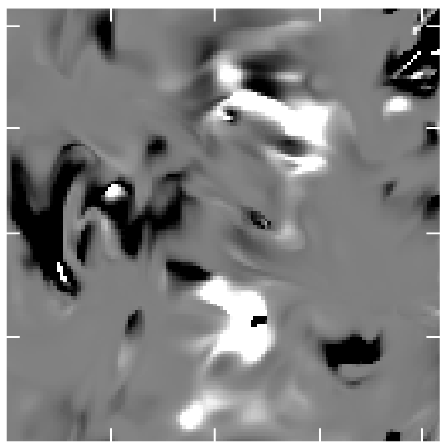}\includegraphics[]{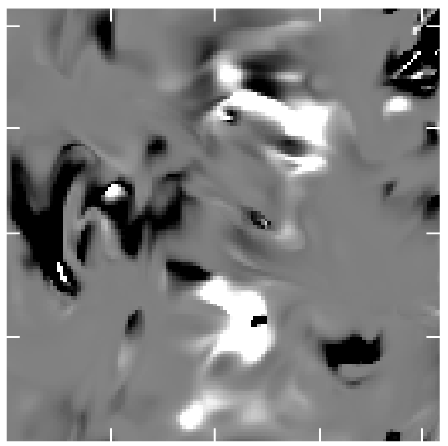}\includegraphics[]{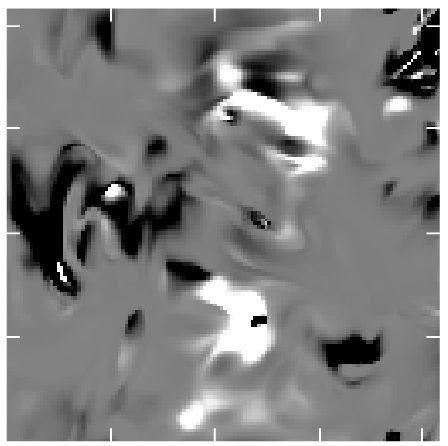}}
   \resizebox{\hsize}{!}{\includegraphics[]{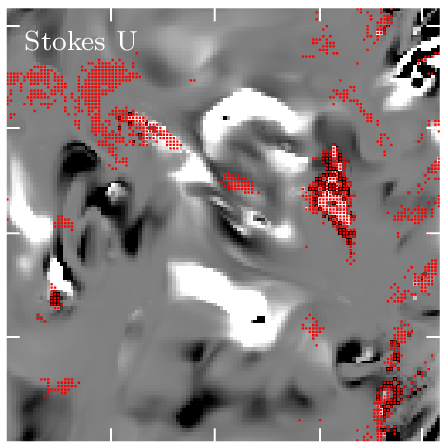}\includegraphics[]{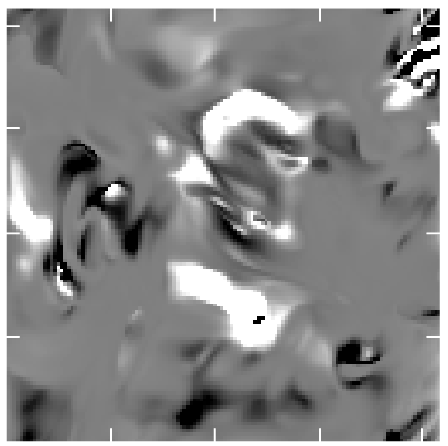}\includegraphics[]{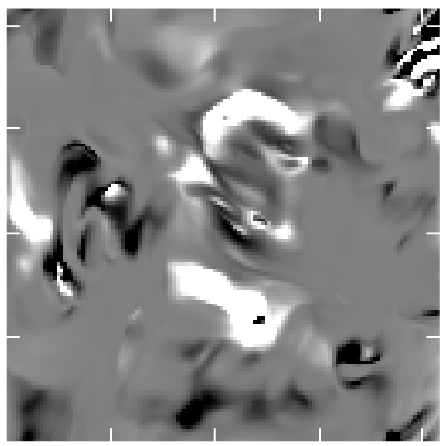}\includegraphics[]{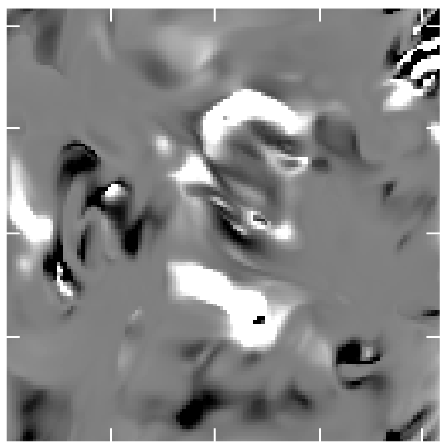}}
   \resizebox{\hsize}{!}{\includegraphics[]{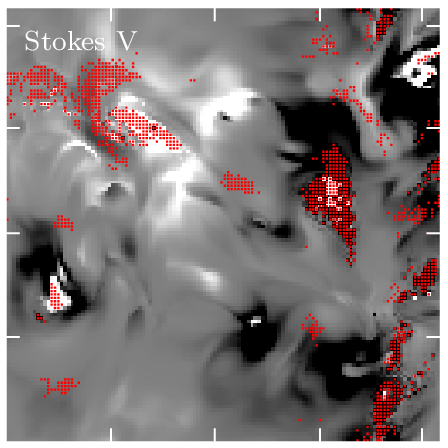}\includegraphics[]{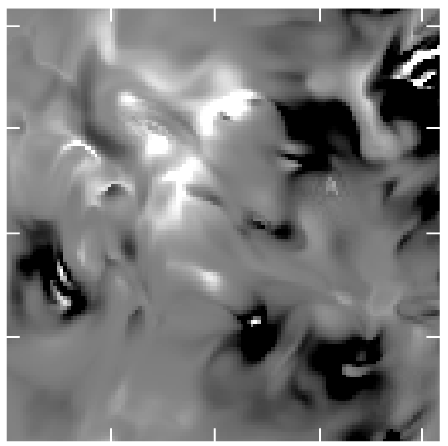}\includegraphics[]{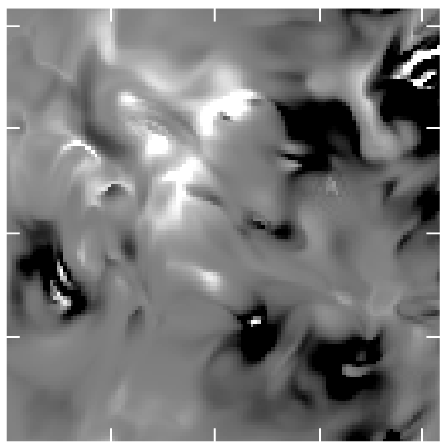}\includegraphics[]{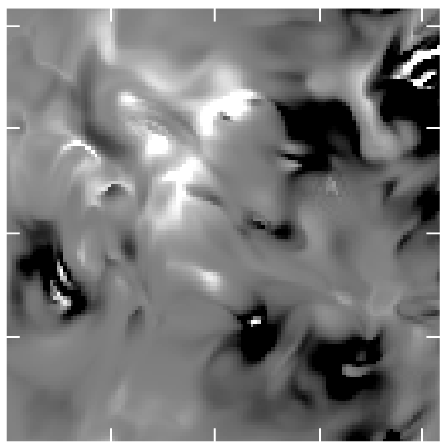}}
\caption{Monochromatic full-Stokes images computed at the core of the
  \ion{Ca}{2}~$\lambda8542$ line. From left to right, the images are computed using the
  DELO-parabolic, DELO-linear, DELO-Bezier and cubic DELO-Bezier,
  respectively. Stokes $I$, $Q$, $U$ and $V$ are represented from top
  to bottom respectively. Regions with artifacts are marked in the
  panels computed using the DELO-par method using small red squares.}  \label{fig:2}
 \end{figure*}

\section{Numerical calculations}\label{sec:numcalc}
We use a modified version of the radiative transfer code \textsc{Nicole} (Socas-Navarro et al. in prep), to test the
numerical accuracy of the DELO-Bezier methods for polarized light. A snapshot from a 3D MHD numerical simulation is used to
carry-out the calculation of synthetic full-stokes profiles. It covers a physical range of $16.6\times8.3\times15.5$ Mm,
extending from the upper convection zone to the lower corona (from 1.5 Mm below to 14 Mm above average optical depth unity at 5000~\AA). The simulation has an average magnetic field strength of 150~G in the photosphere.
This particular snapshot has been used by \citet{2009leenaarts,2012delacruz} so further details on simulations can be found there.

\subsection{The NLTE problem}
As described in \citet{2012delacruz}, the \textsc{Nicole} code solves
the NLTE problem for unpolarized light \citep{1997socas-navarro} and, once the atom population densities are
known, a polarized formal solution is calculated \citep[the
polarization-free approximation, see][]{1996trujillo-bueno}. The
\ion{Ca}{2} atom model used in this work consists of five bound levels
plus a continuum.

Additionally, the \emph{velocity-free} approximation, previously
used to compute the data inversions in \citet{2000socas-navarro,2012delacruz},
is not used in this study. Therefore the population densities are fully
consistent with the strong velocity gradients present in our models.

For consistency, we have implemented an accelerated local lambda operator based in the
unpolarized solution described in Sec.~\ref{sec:bezscal}.

The process of constructing the approximate lambda operator at any
given point $k$, can be idealized by setting the source function to
unity in that point, and zero otherwise. The operator is constructed
using the terms from Equation~(\ref{eq:DELOscalsol}) that remain after this operation
\citep{1987olson}. 

According to Equation~(\ref{eq:tcon}), the average control
point $C$ has an explicit dependency on $C^0_k(S_k)$ and
$C^1_k(S_{k+1})$. Using the
average of the two $C$'s improves stability and
accuracy of the Bezier interpolant, although the lambda operator becomes intrinsically
\emph{non local}. 

A simple recipe for reducing the number of iterations for the NLTE
problem is to make the approximate lambda operator strictly local by
setting $C=C^0_k$. Rare cases of minor overshooting are suppressed by forcing the control point 
$\text{min}(S_k,S_{k+1})\leq C_k \leq \text{max}(S_k,S_{k+1})$. Similar strategies are found in \citet{2010hayek} and \citet{2012holzreuter}.

We used the approximate lambda operator for solving the NLTE problem
in the form:

\begin{equation}
  \Lambda^*_k = r_{\nu\mu}(\alpha_k + \gamma_k),
\end{equation}
where $r_{\nu\mu}$ is the ratio between the line opacity and the total opacity \citep[see][for further details]{1991rybicki}.

Fig.~\ref{fig:2} illustrates the full-Stokes images at the surface of a
$8.3\times8.3$~Mm patch computed at $-10$~m\AA\ from the core of the
\ion{Ca}{2}~$\lambda8542$ line. Columns of panels (from left to right)
show the monochromatic images obtained using DELO-parabolic,
DELO-linear, quadratic DELO-Bezier and cubic DELO-Bezier, respectively.
We used identical NLTE level populations to compute the polarized
formal solution with each algorithm. Thus, the differences between
panels in different columns reflect the numerical properties (convergence
and stability) of the methods compared.

The DELO-parabolic method seems to work in most of the pixels, but it fails
notoriously to produce an accurate solution at certain areas which have
been indicated on the panels using red markers. The artifacts here can
be quite large, and some pixels show even negative values in Stokes~$I$.

The maximum difference in Stokes $I$ between DELO-linear and
DELO-Beziers at this wavelength is less than $1\%$ of the
continuum intensity. This is an expected result, given that the
vertical grid spacing of 3D snapshots is so fine that even a linear scheme
produces an accurate solution.

This example only shows the stability of the higher-order Bezier
methods, but it hardly shows any advantage
over a linear scheme.

In Sec.~\ref{sec:numac} we test the converges and stability of these methods using
1D models where strong gradients in the magnetic-field and line-of-sight (l.o.s) velocity are
introduced.

\subsection{Numerical accuracy}\label{sec:numac}
To assess the accuracy of the formal solvers, we have created
four atmospheric models using a fine grid of depth-points with 198
points-per-decade, equidistantly placed from
$\log \tau_{500}=-6.9$ to $\log \tau_{500}=2$.

The four atmospheric models share the same temperature,
electron pressure, and micro-turbulence from the VALC-III model \citep{1981vernazza}.

\begin{figure}
  \centering
\resizebox{\hsize}{!}{\includegraphics[]{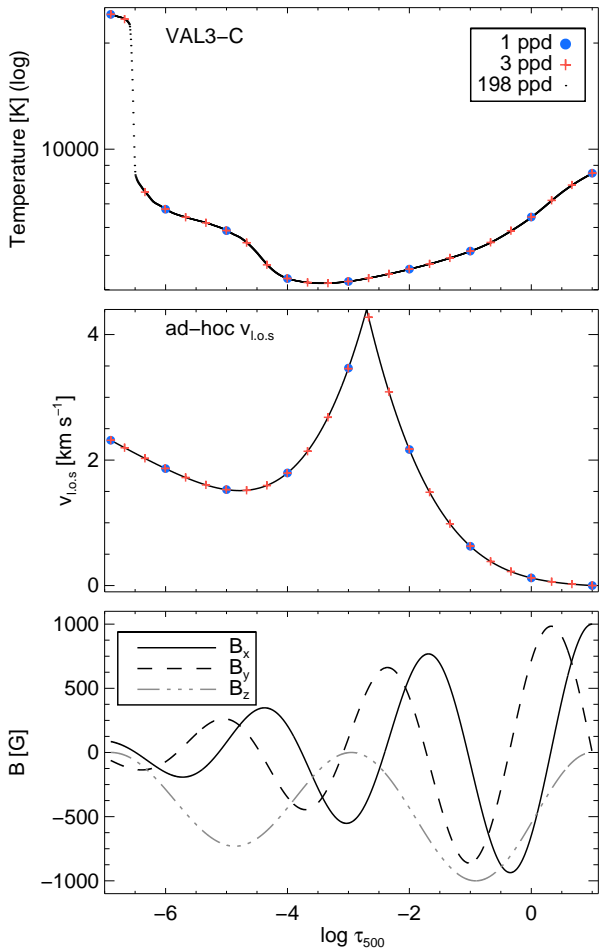}}
\caption{Physical quantities from our prescribed 1D stellar model. The
top panel shows the temperature stratification as a function of
optical depth. The middle panel shows an ad-hoc profile used to define
 $v_{l.o.s}$ in our calculations. The bottom panel
illustrates the three components of the magnetic field. To
illustrate the effects of having a poor sampling in the vertical
dimension, we have marked in the first two panels, the equivalent
model assuming 1 point-per-decade (blue circles), 3 points-per-decade
(red crosses) and 198 points-per-decade (black dots).  Note that the $z$-axis is parallel to the line-of-sight.}\label{fig:3}
\end{figure}

We have created a complicated ad-hoc magnetic field
vector and l.o.s velocity component, which are illustrated in Fig.~\ref{fig:3},
along with the VALC-III temperature. To create a 
demanding topology, the horizontal component of the magnetic-field follows
a spiral centered on the line-of-sight. 

Considering the quantities described above, our set of four models is
summarized as follows:
\begin{enumerate}
  \item The first case is a model with constant l.o.s velocity
    ($v_{l.o.s}=0$~km s$^{-1}$) and constant magnetic-field
    $\mathbi{B} = (600,-700,800)$~G for all depth-points.

   \item In the second case, the constant
     line-of-sight velocity is replaced with our ad-hoc profile.

    \item The third case has constant velocity but the complicated
      magnetic-field vector described in Fig.~\ref{fig:3}.

    \item Complex velocity and magnetic-field profiles are used
      simultaneously in the fourth case.
\end{enumerate}

For each of these four cases, we have computed the absorption matrix
and the source function in the original grid of depth-points for the
\ion{Ca}{2}~$\lambda8542$ and the
\ion{Fe}{1}~$\lambda6302$ lines. These two lines have been extensively used as
atmospheric diagnostics in solar applications. The former is
sensitive to a vast range of height including photospheric and
chromospheric response, although it has an relatively low
Land\'e factor  ($g_\text{eff}=1.1$). The latter is formed in the solar photosphere,
but it has a much higher sensitivity to the magnetic-field
($g_\text{eff}=2.5$).

Our test consists of computing the formal solution to the polarized
RTE using a pre-computed absorption matrix and source function. To study
the converge properties of the integrators we simply drop intermediate
depth-points from the original grid. Thus we separate the level population
calculations from the RTE solution. We also noticed that our level population
calculations for \ion{Ca}{2} start deviating noticeably from the correct
solution when the number of depth-points is below 10 points-per-decade.

\begin{figure}

\resizebox{\hsize}{!}{\includegraphics[]{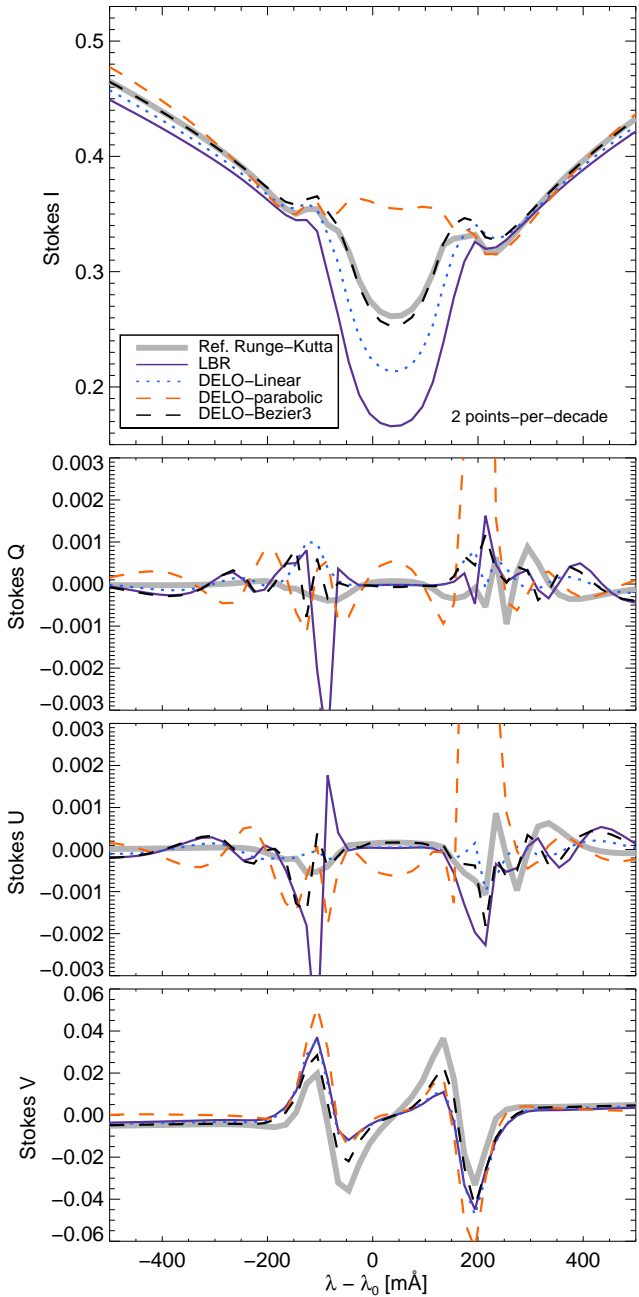}}
\caption{Emerging intensity vector at the \ion{Ca}{2}~$\lambda8542$ line. The
  profiles are computed using our fourth model which contains gradients
  in the magnetic-field and line-of-sight-velocity. The model here has
  a vertical resolution of of 2 points-per-decade. The
  reference profile is computed using a Runge-Kutta solver using the
  original grid of 198 points-per-decade. For readability, only the cubic DELO-Bezier solution is plotted.}\label{fig:4}
\end{figure}
As a  reference here we use the emerging Stokes vector computed on the
original grid of 198 points-per-decade, using an adaptive fourth-order
Runge-Kutta solver \citep[see][]{1976landi}, which ensures that the
precision of $10^{-13}$ as achieved at every step.

Fig.~\ref{fig:4} shows a set of Stokes~$I$, $Q$, $U$ and $V$ profiles computed for the
complex of our ad-hoc models (model 4) using a 2 points-per-decade grid. The
reference grey profile has been computed
using the Runge-Kutta method and the original dense grid of
depth-points. The main discrepancies occur at the chromospheric NLTE
core of the \ion{Ca}{2}~$\lambda8542$ line. At those wavelengths, the
monochromatic depth-scale presents large irregular jumps in
optical-depth.

For example, in Stokes~$I$ the LBR and DELO-linear solutions present an excessively strong
line core, whereas the DELO-parabolic method produces a line core in
emission. The DELO-Bezier profiles tightly trace the reference profile at all
wavelengths, and only a noticeable  small deviation is present at the
very inner core of the line.

\begin{figure*}[t]
  \centering
\resizebox{\hsize}{!}{\includegraphics[trim = 0cm 0.65cm 0cm 0cm,clip]{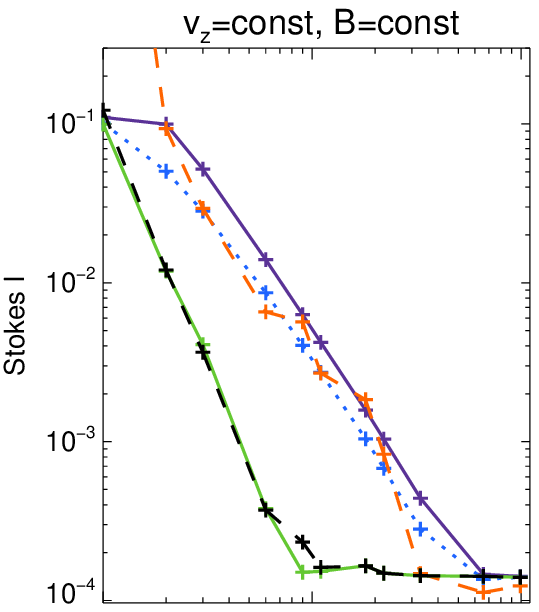}\includegraphics[trim = 0.5cm 0.65cm 0cm 0cm,clip]{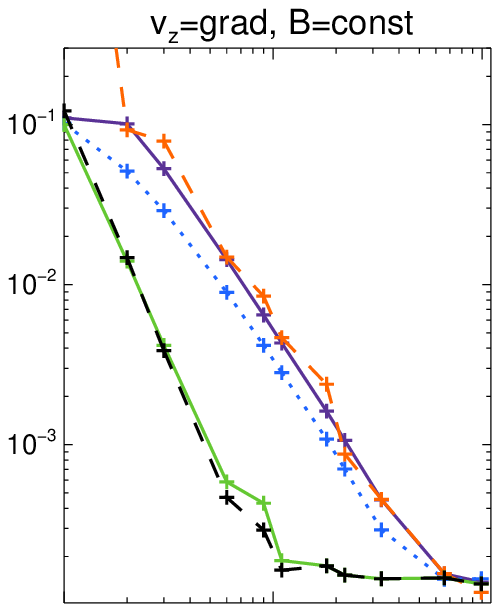}\includegraphics[trim =0.5cm 0.65cm 0cm 0cm, clip]{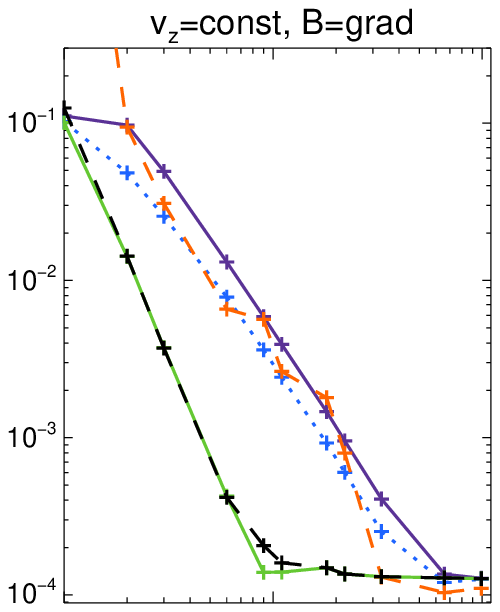}\includegraphics[trim = 0.5cm 0.65cm 0cm 0cm,clip]{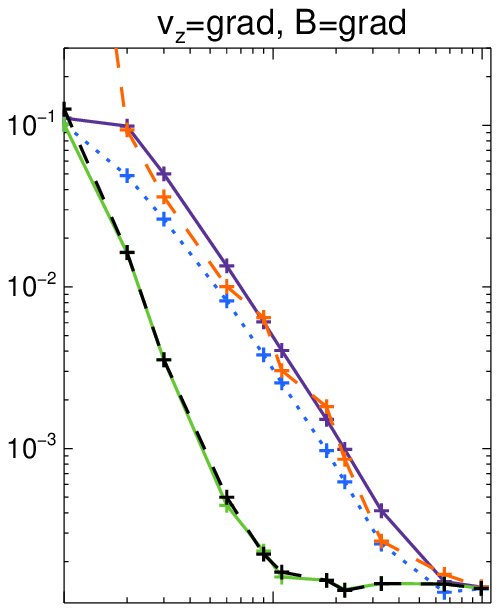}}
\resizebox{\hsize}{!}{\includegraphics[trim = 0cm 0.65cm 0cm 0.5cm, clip]{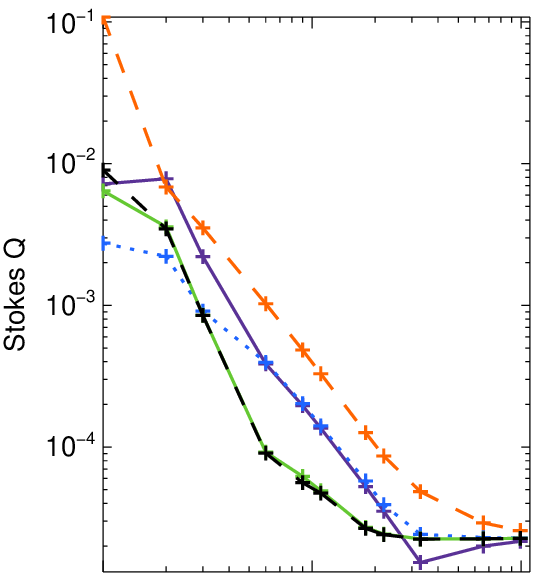}\includegraphics[trim = 0.5cm 0.65cm 0cm 0.5cm, clip]{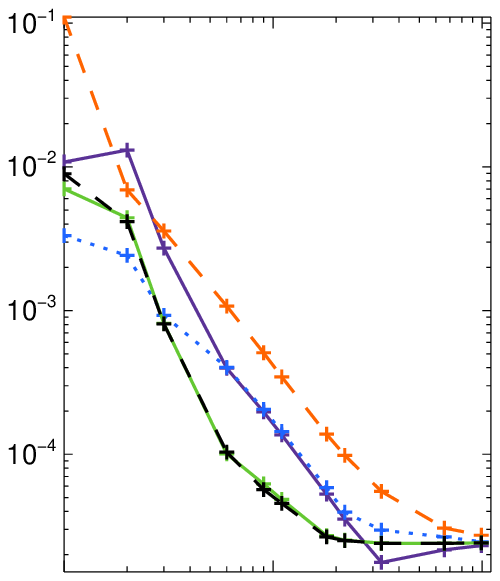}\includegraphics[trim = 0.5cm 0.65cm 0cm 0.5cm, clip]{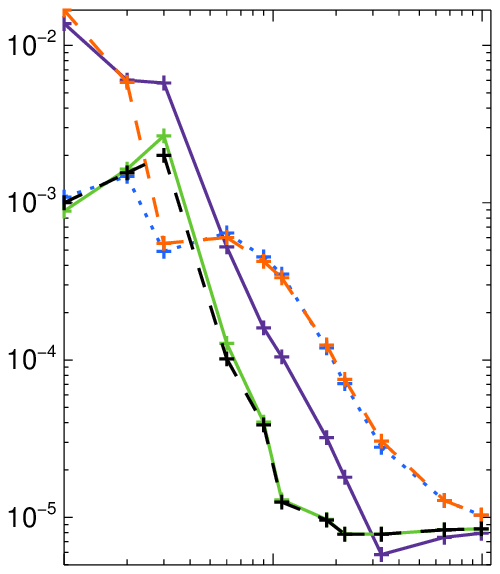}\includegraphics[trim = 0.5cm 0.65cm 0cm 0.5cm, clip]{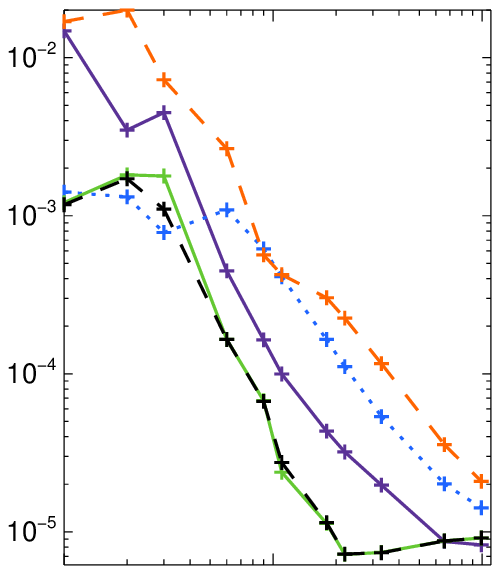}}
\resizebox{\hsize}{!}{\includegraphics[trim = 0cm 0.65cm 0cm 0.5cm, clip]{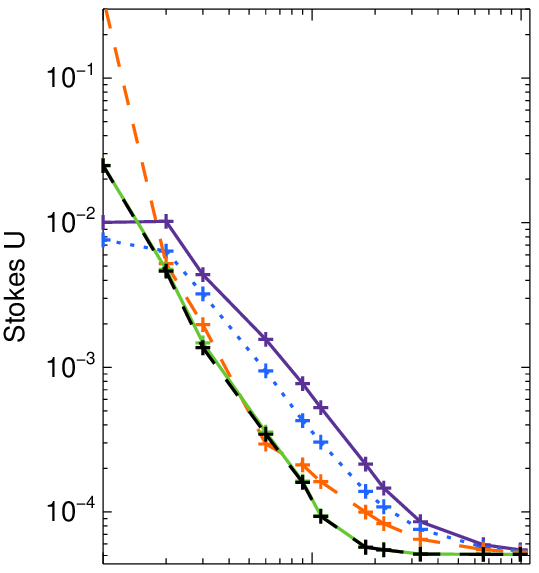}\includegraphics[trim = 0.5cm 0.65cm 0cm 0.5cm, clip]{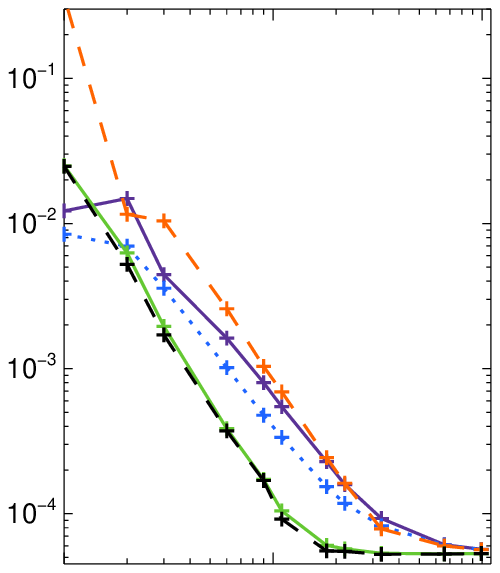}\includegraphics[trim = 0.5cm 0.65cm 0cm 0.5cm, clip]{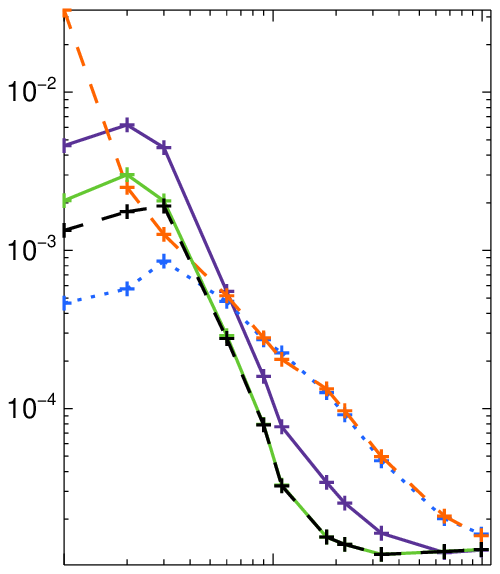}\includegraphics[trim = 0.5cm 0.65cm 0cm 0.5cm, clip]{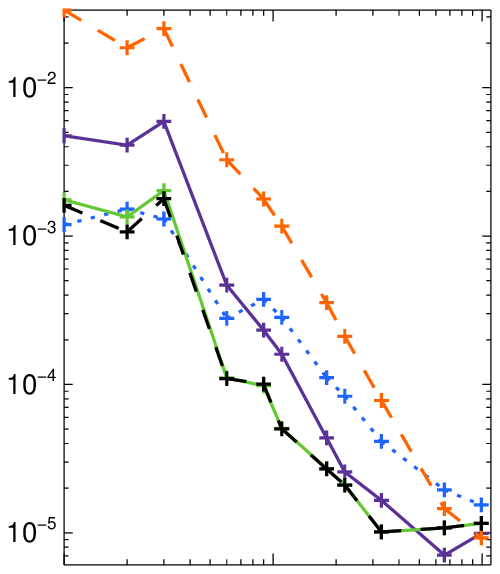}}
\resizebox{\hsize}{!}{\includegraphics[trim = 0cm 0cm 0cm 0.5cm, clip]{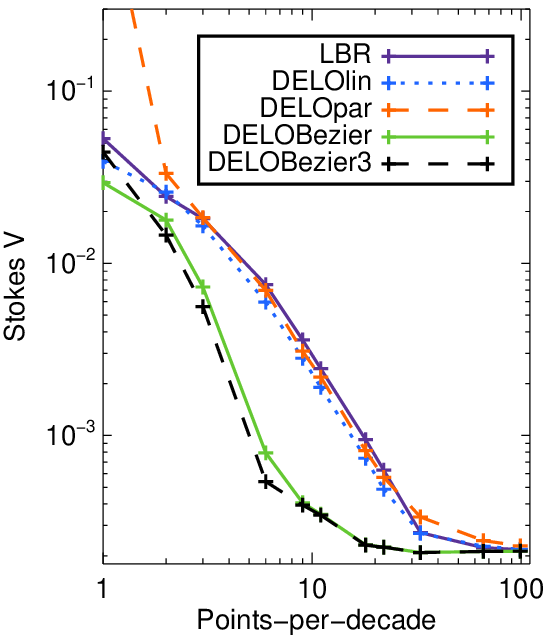}\includegraphics[trim = 0.5cm 0cm 0cm 0.5cm, clip]{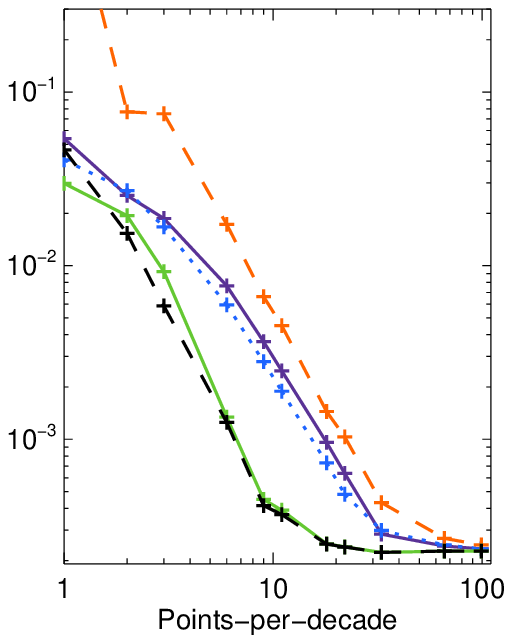}\includegraphics[trim = 0.5cm 0cm 0cm 0.5cm, clip]{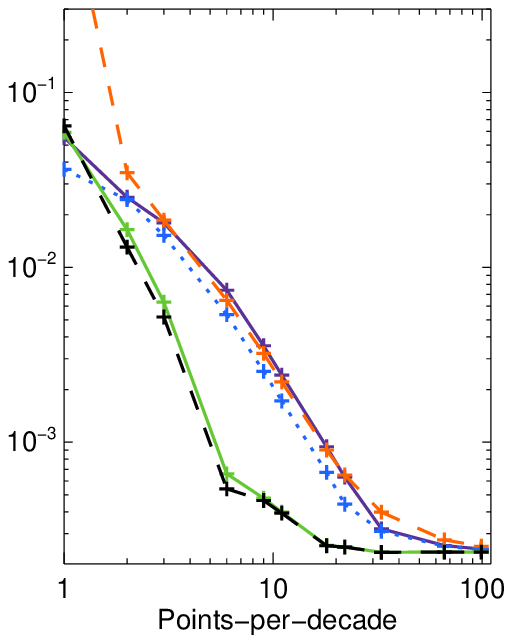}\includegraphics[trim = 0.5cm 0cm 0cm 0.5cm, clip]{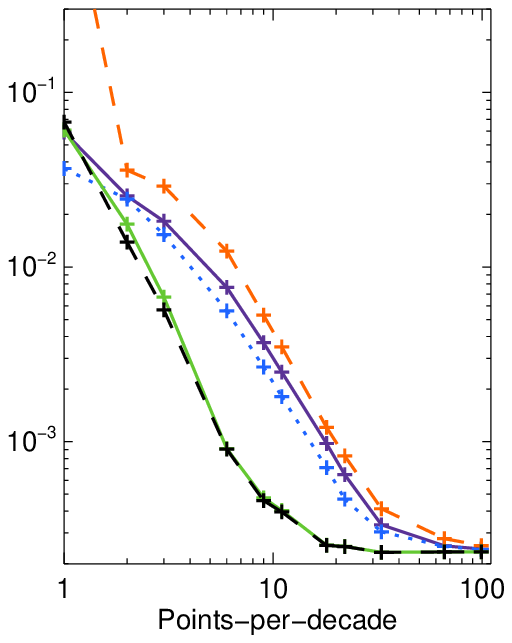}}

\caption{Maximum error for the
  \ion{Ca}{2}~$\lambda8542$ line as a function of the number of
  points-per-decade (at all wavelengths).  From top to bottom, the
  panels show the errors for  Stokes
  $I$, $Q$, $U$ and $V$, respectively. From left to right, each column
corresponds to each of the four models used in our calculations. Each
colored line corresponds to one of the formal solutions used in our
calculations: LBR (solid purple), DELO-linear (dotted blue),
DELO-parabolic (dashed orange), quadratic DELO-Bezier (solid green) and
cubic DELO-Bezier (dashed black). }\label{fig:5}
\end{figure*}

\begin{figure*}[t]
  \centering
\resizebox{\hsize}{!}{\includegraphics[trim = 0cm 0.65cm 0cm 0cm,clip]{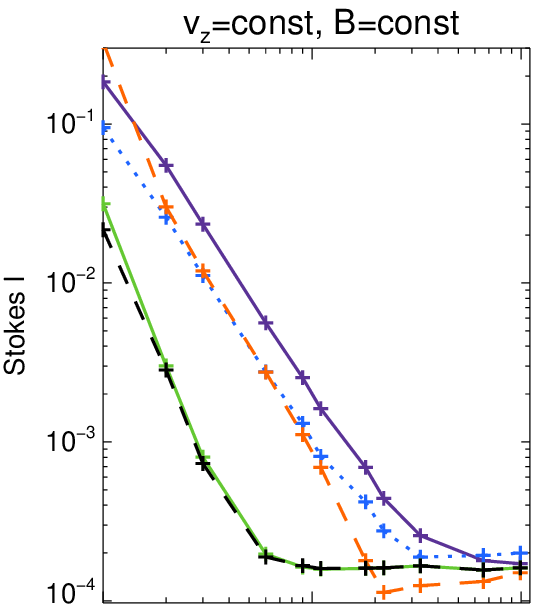}\includegraphics[trim = 0.5cm 0.65cm 0cm 0cm,clip]{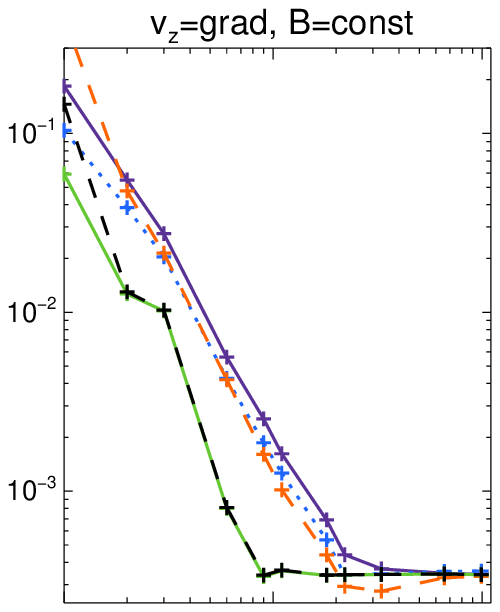}\includegraphics[trim =0.5cm 0.65cm 0cm 0cm, clip]{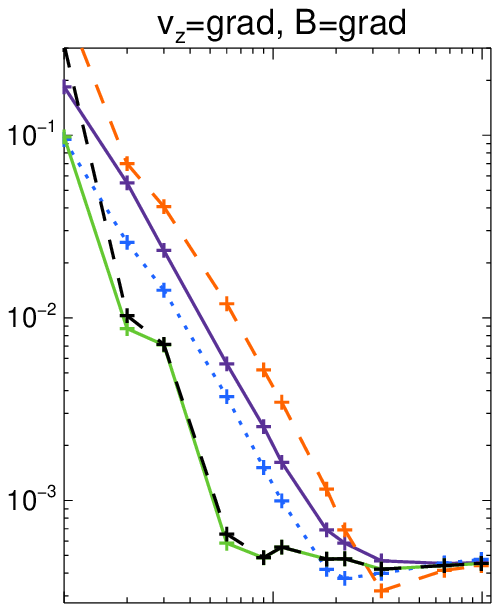}\includegraphics[trim = 0.5cm 0.65cm 0cm 0cm,clip]{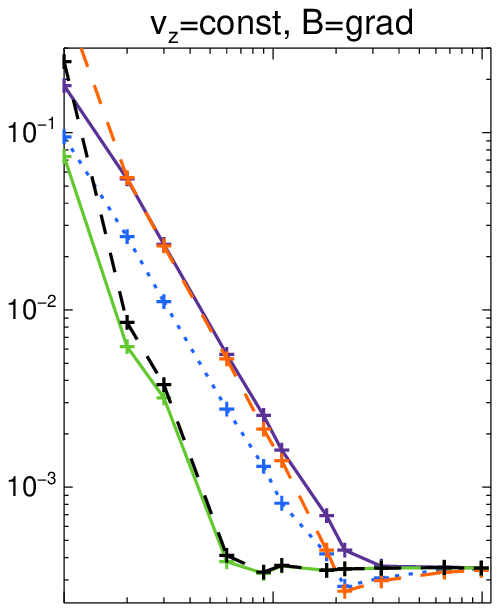}}
\resizebox{\hsize}{!}{\includegraphics[trim = 0cm 0.65cm 0cm 0.5cm, clip]{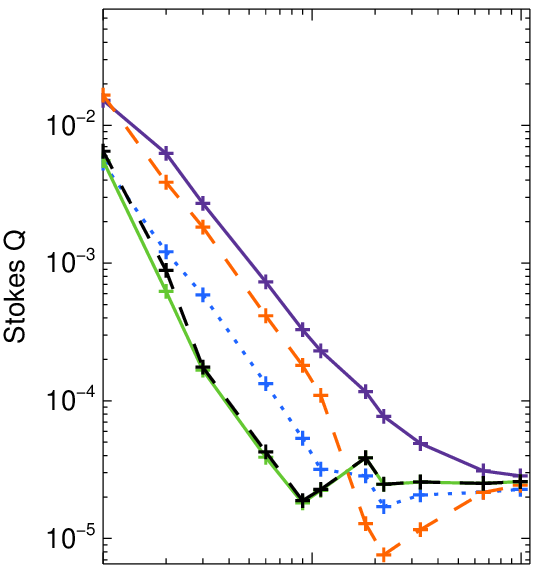}\includegraphics[trim = 0.5cm 0.65cm 0cm 0.5cm, clip]{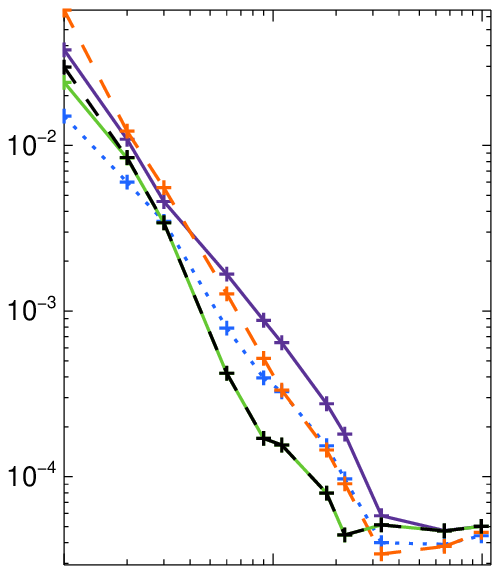}\includegraphics[trim = 0.5cm 0.65cm 0cm 0.5cm, clip]{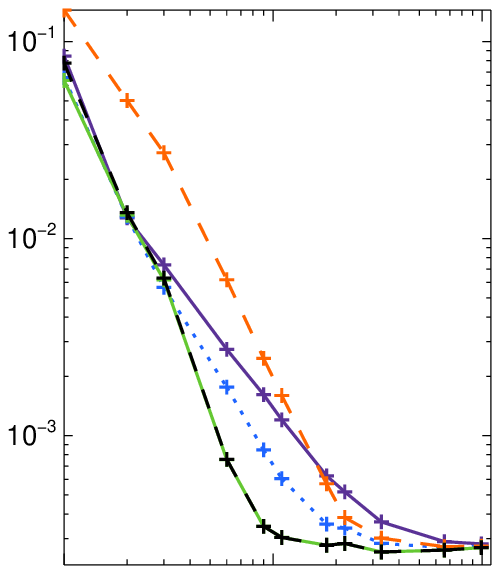}\includegraphics[trim = 0.5cm 0.65cm 0cm 0.5cm, clip]{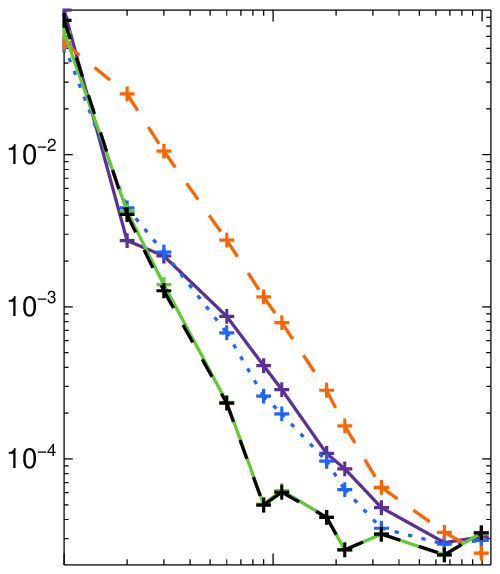}}
\resizebox{\hsize}{!}{\includegraphics[trim = 0cm 0.65cm 0cm 0.5cm,clip]{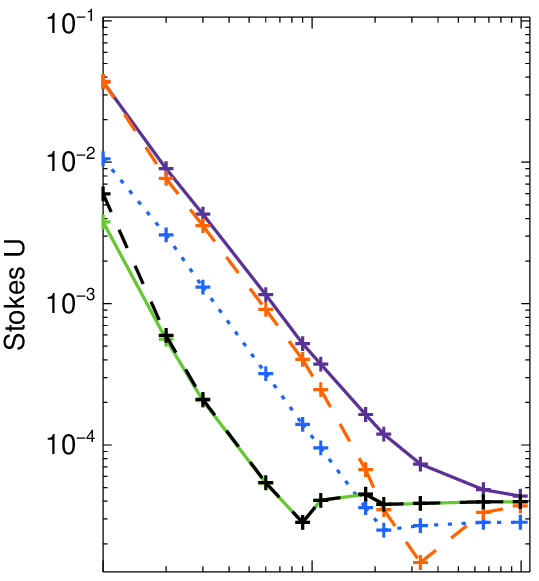}\includegraphics[trim = 0.5cm 0.65cm 0cm 0.5cm,clip]{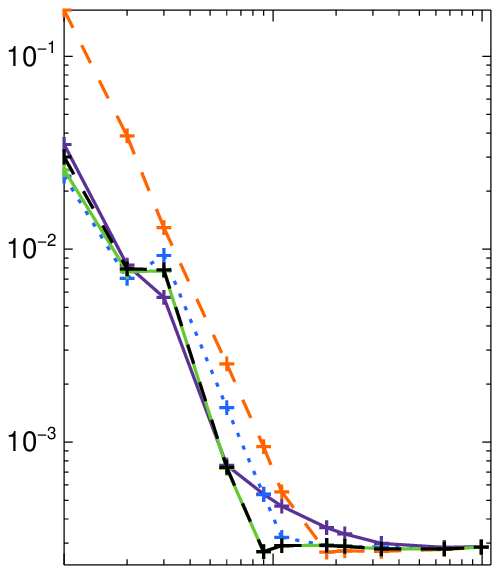}\includegraphics[trim = 0.5cm 0.65cm 0cm 0.5cm,clip]{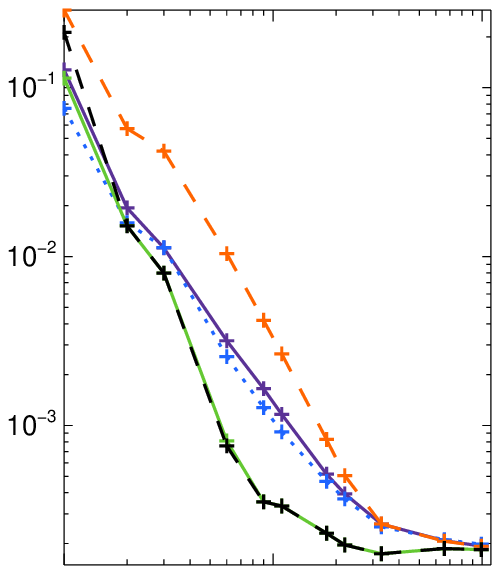}\includegraphics[trim = 0.5cm 0.65cm 0cm 0.5cm, clip]{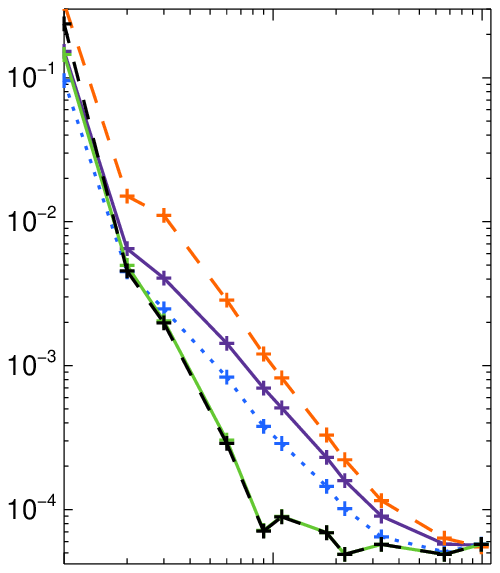}}
\resizebox{\hsize}{!}{\includegraphics[trim = 0cm 0cm 0cm 0.5cm, clip]{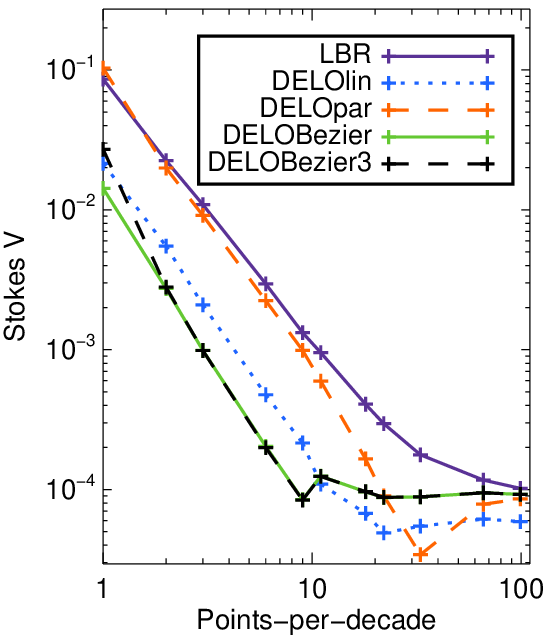}\includegraphics[trim = 0.5cm 0cm 0cm 0.5cm, clip]{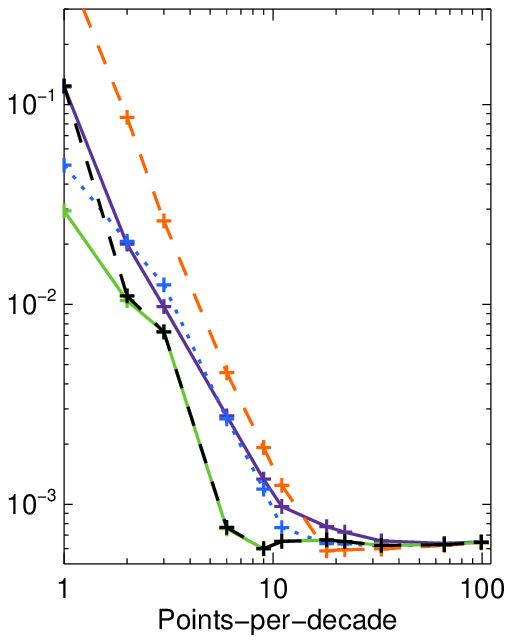}\includegraphics[trim = 0.5cm 0cm 0cm 0.5cm, clip]{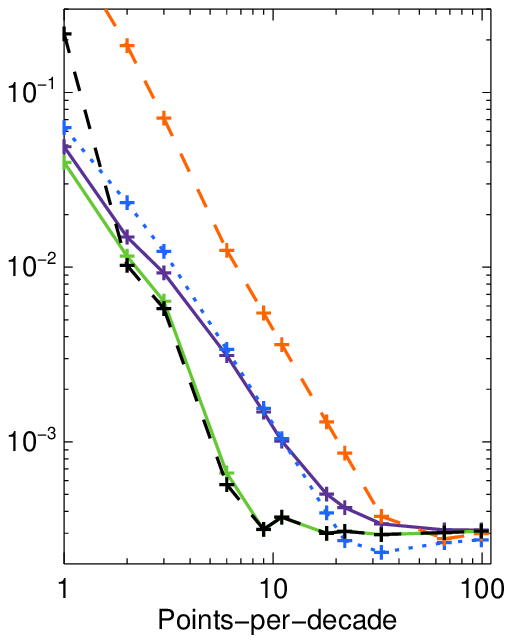}\includegraphics[trim = 0.5cm 0cm 0cm 0.5cm, clip]{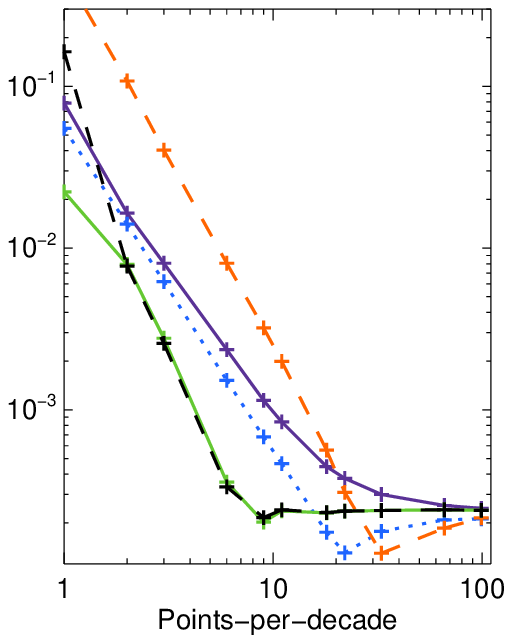}}

\caption{Maximum error for the
  \ion{Fe}{1}~$\lambda6302$ line as a function of the number of
  points-per-decade (at all wavelengths).  From top to bottom, the
  panels show the errors for  Stokes
  $I$, $Q$, $U$ and $V$, respectively. From left to right, each column
corresponds to each of the four models used in our calculations. Each
colored line corresponds to one of the formal solutions used in our
calculations: LBR (solid purple), DELO-linear (dotted blue),
DELO-parabolic (dashed orange), quadratic DELO-Bezier (solid green) and
cubic DELO-Bezier (dashed black). }\label{fig:6}

\end{figure*}

The results of our calculations are summarized in Fig.~\ref{fig:5}
and Fig.~\ref{fig:6}, for the \ion{Ca}{2}~$\lambda8542$ line and the
\ion{Fe}{1}~$\lambda6302$ line respectively.  The line profiles are compared with the reference
		case, and the largest discrepancy over the entire
		profile (in absolute value) is plotted for each
		integration method as a function of the number of
		depth-points per decade.

This error measurement is sensitive to outliers at a single
wavelength. Therefore, the curves in Fig.~\ref{fig:5} and
Fig.~\ref{fig:6} may not be monotonous functions of the number of
grid points. Still, we prefer this metric as it gives a robust
estimate of accuracy, convergence speed and stability for our solvers.

Ideally, we would expect that any discrepancies shown by all the
methods considered here, would disappear when the density of
depth-points is large enough (all methods should converge to the same
accurate solution).  The errors
are expected to be large when the density of depth-points is low,
especially for the highest order methods, but these should achieve a
more accurate solution than the lower order methods when the density
of depth-points is relatively high.

In Fig.~\ref{fig:5}, our results show that at the absolute minimum
of 1 points-per-decade, it is hard to beat the DELO-linear
solution, which keep the errors under control despite the strong
gradients and poorly-sampled stellar atmosphere. However, the
DELO-Bezier solutions provide almost as accurate results except in
Stokes~$Q$ where the error is half an order of magnitude higher.

The situation changes drastically, as a few more depth-points are
included in the calculations. Between five and thirty points-per-decade,
the DELO-Bezier solutions clearly out-perform all the other solutions
for the four scenarios that we have prepared. Most radiative transfer
computations are carried out within this range of grid densities. This
also matches hydrodynamic grid density typically used for environments
where radiation carries important fraction of energy.

The DELO-parabolic needs a large number of depth-points to match the
performance of all the other methods, as was expected.

The calculations for the \ion{Fe}{1}~$\lambda6302$ line, show a
smoother behavior than in the previous case and all cases seem to
reach a stable level of accuracy at approximately 30 points-per-decade.
The inclusion of more points above this value, seems to
increase the accuracy marginally and very inefficiently.

The DELO-Bezier demonstrated fastest convergence reaching below 10\,\%~
for the least dense grids in Stokes~$I$. This conclusion is true for all our models.
The accuracy of all methods is comparable as they all reach the same
level of precision above 30 points-per-decade.

For both spectral lines, the performance of the LBR method is close to
both the DELO-linear and DELO-parabolic methods, and it only seems to be
slightly less accurate when strong gradients in the line-of-sight
velocity are present.

The accuracy achieved by the DELO-Bezier methods is
almost identical. The quadratic solution marginally
outperforms the cubic one when the number of
depth-points per decade is 1 or 2, whereas the cubic version is slightly 
better at the \ion{Ca}{2}~$\lambda8542$ line, within the range $2<\text{N}_\text{depth}<10$ points-per-decade. 

Computationally, the LBR and DELO-Bezier solutions
are similarly demanding given that  accurate derivatives of the source
function and of the absorption matrix are required.
In comparison, the DELO-linear and DELO-parabolic solutions are almost
two times faster, because no derivatives are required.

Note also that, in the first and last intervals of the grid, only one control-point can be computed. 
Therefore, especial care must be taken when the cubic method is used. The simplest solution is 
to use the quadratic method in those two intervals, given that only one control point is strictly necessary.


\section{Discussion and conclusions}\label{sec:discussion}
In this work we present a new scheme to integrate the polarized
radiative transfer equation using Bezier splines. The solvers are a
generalization of the advanced strategy
proposed by \citet{2003auer} for the unpolarized case. Our
second-order and third-order integration methods preserve the stability of the
DELO-linear solution with the benefits of faster convergence and higher order accuracy.

The advantages of the new integration scheme are illustrated by
comparison with other methods commonly used to solve the polarized
RTE. We show that the new DELO-Bezier schemes outperform all other
algorithms when the density of depth-points is low and the structure
of the medium includes steep gradients.

The new methods will be beneficial for various application from reconstructing
atmospheric structure(s) using observational data (data inversion in solar physics,
Magnetic Doppler Imaging of stars) to radiative magneto-hydrodynamics. E.g., in case
of solar data inversion the new solvers will allow using less denser depth grid
improving the stability of inversion, which is particularly important for strong
lines, like the \ion{Ca}{2} H,K and the IR triplet. Given the vast formation
range of these lines (larger than $1000$~km), it is hard to find an optimal
grid of depth points at all frequencies in the line.

\acknowledgments
The Institute of Theoretical Astrophysics
(University of Oslo) is gratefully acknowledged for providing a
snapshot from a 3D MHD chromospheric simulation. JdlCR gratefully
acknowledges valuable scientific discussion with Andr\'es
Asensio-Ramos during the implementation of the approximate lambda
operator. The authors of this work are grateful to an anonymous
referee, whose suggestions and comments helped to improve the text.
\clearpage

\appendix

\section{Bezier integral of the opacity}\label{sec:apen1}
In many radiative transfer applications, a conversion of the
depth-scale is desirable (i.e, from $z$ to $\log\tau_{500}$). To carry
out this conversion, the opacity ($\eta_I \equiv \eta$) is integrated along the
depth-scale. When the grid of depth-points is sufficiently dense, a
trapezoidal integration would suffice. This wide-spread method becomes
inaccurate for coarse depth-scale grids.

We advice to approximate the opacity with a quadratic Bezier spline (Eq.~\ref{eq:bezier}), and integrate
analytically this function. Once again, the integral can be expressed
as a set of coefficients that multiply the values of the opacity that
define the integration interval:
\begin{equation}
\begin{split}
  d\tau_{k}(\nu) = \tau_{k+1} - \tau_{k}=&  (x_{k+1}-x_k)\int^1_0 \left [\eta_k(\nu)*(1-u)^2
    + \eta_{k+1}(\nu)*u^2+2u(1-u)C \right] du = \\
  = &
  \frac{x_{k+1}-x_{k}}{3} \left (\eta_{k}(\nu) +\eta_{k+1}(\nu) + C \right).
\end{split}
\end{equation}

\section{Taylor expansion of the Bezier integral coefficients}\label{sec:apen2}
In \S~\ref{sec:bezscal} we advise to use a Taylor expansion of the
exponential term in Equation~(\ref{eq:DELOsol}), when $\delta_k$ is
small:
\begin{equation*}
e^{-\delta_ku} \approx 1 - \delta_ku + \frac{\delta_k^2u^2}{2} - \frac{\delta_k^3u^3}{6}.
\end{equation*}

The resulting integration coefficients for the quadratic Bezier
interpolant are:
\begin{eqnarray*}
  \alpha_k &=&    \frac{\delta_k}{3} - \frac{\delta_k^2}{12} + \frac{\delta_k^3}{60},\\
  \beta_k &=&      \frac{\delta_k}{3} - \frac{\delta_k^2}{4} + \frac{\delta_k^3}{10},\\
  \gamma_k &=& \frac{\delta_k}{3} - \frac{\delta_k^2}{6} +
  \frac{\delta_k^3}{20}.
\end{eqnarray*}

The integration coefficients for the cubic Bezier interpolant are:
\begin{eqnarray*}
\hat{\alpha_k} &=& \frac{\delta_k}{4} - \frac{\delta_k^2}{20} + \frac{\delta_k^3}{120} - \frac{\delta_k^4}{840},\\
\hat{\beta_k} &=& \frac{\delta_k}{4} - \frac{\delta_k^2}{5} + \frac{\delta_k^3}{12} -  \frac{\delta_k^4}{42},\\
\hat{\gamma_k} &=& \frac{\delta_k}{4} - \frac{\delta_k^2}{10} + \frac{\delta_k^3}{40} - \frac{\delta_k^4}{210},\\
\hat{\epsilon_k} &=& \frac{\delta_k} {4} - \frac{3\delta_k^2}{20} + \frac{\delta_k^3}{20} - \frac{\delta_k^4}{84}.
\end{eqnarray*}


\end{document}